\documentstyle[aps,prl,multicol,amssymb]{revtex}
\title{Violations of local realism by two entangled quNits}

\author{Daniel Collins$^{1,2}$ and Sandu Popescu$^{1,2}$}

\address{ $^1$H.H. Wills Physics Laboratory,
University of Bristol, Tyndall Avenue, Bristol BS8 1TL, UK
\\ $^2$BRIMS, Hewlett-Packard Laboratories,
 Stoke Gifford, Bristol BS12 6QZ, UK}
\date{26 June 2001}
\begin{document}
\maketitle
\begin{abstract}

Results obtained in two recent papers, \cite{Kaszlikowski} and
\cite{Durt} seem to indicate that the nonlocal character of the
correlations between the outcomes of measurements performed on
entangled systems separated in space is not robust in the presence
of noise. This is surprising, since {\it entanglement} itself is
robust. Here we revisit this problem and argue that the class of
gedanken-experiments  considered in \cite{Kaszlikowski} and
\cite{Durt}  is too restrictive. By considering a more general
class, involving {\it sequences} of measurements, we prove that
the nonlocal correlations are in fact robust.

\end{abstract}


\newcommand{\tr}{\mbox{Tr} }
\newcommand{\ket}[1]{\left | #1 \right \rangle}
\newcommand{\bra}[1]{\left \langle #1 \right |}
\newcommand{\amp}[2]{\left \langle #1 \left | #2 \right. \right \rangle}
\newcommand{\proj}[1]{\ket{#1} \! \bra{#1}}
\newcommand{\ave}[1]{\left \langle #1 \right \rangle}
\newcommand{\superop}{{\cal E}}
\newcommand{\unity}{\mbox{\bf I}}
\newcommand{\hilbert}{{\cal H}}
\newcommand{\relent}[2]{S \left ( #1 || #2 \right )}
\newcommand{\banner}[1]{\bigskip \noindent {\bf #1} \medskip}
\newcommand{\I}{{\mathbf I}}
\newcommand{\R}{{\mathbf R}}
\renewcommand{\S}{{\mathbf S}}
\newcommand{\up}{\uparrow}
\newcommand{\down}{\downarrow}

 \begin{multicols}{2}


In his famous paper \cite{Bell}, J. Bell showed that quantum
mechanics predicts nonlocal correlations between measurement
outcomes at spatially separated regions in a certain experiment.
By nonlocal correlations we mean correlations which cannot be
explained by any local hidden variable model (LHV).  During the
last few years other aspects of nonlocality, in addition to
generating nonlocal correlations have been discovered.  For
example, the ability of quantum states to teleport\cite{teleport},
to super-dense code\cite{superdense}, and to reduce the number of
classical bits required to perform certain communication tasks (in
the so called ``communication complexity''
scenario)\cite{complexity}. Further, nonlocality appears to be at
the heart of quantum computation\cite{qcomp} and its ability to
perform certain computations exponentially faster than any
classical device.

Two recent papers \cite{Kaszlikowski} and \cite{Durt}  have
studied the question of robustness of nonlocal correlations.
Results in \cite{Kaszlikowski} and \cite{Durt} seem to indicate a
very surprising result. Namely, it appears that in a certain sense
(which we will define more precisely later), quantum nonlocal
correlations are not very robust. Here we would like to argue that
nonlocal correlations are actually very robust. While we do not
disagree with the specific results found in \cite{Kaszlikowski}
and \cite{Durt}, we show that the class of gedanken experiments
they have considered (though very interesting in itself) is in
fact quite limited and not sensitive enough. We present a
different class of experiments which shows that nonlocal
correlations are robust.

The authors of \cite{Kaszlikowski} and \cite{Durt} have considered
two quantum particles, each living in an $N$ dimensional Hilbert
space, which are in the maximally entangled state mixed with
random noise. ie. states of the form
\begin{equation}
\rho_N (F_N) =  (1 - F_N) \ket{ \Psi_N}_{AB}\bra{ \Psi_N } + F_N
\frac{1}{N^2} \hat{I}_{N \times N},\label{state}
\end{equation}
where
\begin{equation}
\ket{ \Psi_N}_{AB} = \frac{1}{\sqrt{N}} \sum_{m=1}^{N} \ket{m}_A
\ket{m}_B,
\end{equation}
$F_N$ is a constant $0\le F_N\le 1$ which describes the fraction
of noise and $\hat{I}_{N \times N}$ is the identity matrix.  They
have asked, ``what is the maximum fraction of noise, $F_N$, which
can be added to the maximally entangled state so that the state
still generates nonlocal correlations?''

It is useful here to make a clear distinction between two
different issues which are relevant for our discussion. The first
is the issue of {\it entanglement} or {\it non-separability}. A
quantum state is {\it separable} if it can be written as
\begin{equation}
\rho_{AB} = \sum_i p_i \rho_A^i \rho_B^i,
\end{equation}
and it is non-separable otherwise.

It has been shown \cite{Zyczkowski}, \cite{Vidal} and
\cite{Braunstein} that if too much noise is added to the maximally
entangled state, the state ceases to be entangled. Obviously, at
this moment the quantum state ceases to have any nonlocal aspects
whatsoever.

The other issue is whether or not the results of all possible
measurements performed on the state can be explained by a local
hidden variable model. If they cannot we say, following Bell, that
the state generates nonlocal correlations (sometimes this is
called a ``violation of local realism").

It is clear that when there is so much noise that the state
becomes separable, the state cannot generate any nonlocal
correlations. It is however possible that the state ceases to
generate nonlocal correlations at smaller levels of noise, i.e.
while it is still entangled. Indeed, it is not known if every
entangled (mixed) state generates nonlocal correlations or not -
this is one of the most important issues in quantum nonlocality.

It appears from the results of \cite{Kaszlikowski} and \cite{Durt}
that the nonlocal correlations are not robust, meaning that for
fractions of noise greater than $F_N \approx 0.33$ none of the
states $\rho(F_N)$ produce nonlocal correlations. This is very
surprising since the entanglement property of the maximally
entangled states is robust -   for any fraction of noise, when the
dimensionality of the systems is large enough (how large depending
on the fraction of noise), the states of form (\ref{state}) are
entangled. Furthermore, these mixed entangled states exhibit most
other nonlocality aspects - for example they can be used for
teleportation, super-dense coding, and can be purified to yield
singlets. So it would be quite strange if they couldn't also
generate nonlocal correlations.

We shall show that nonlocal correlations are, similar to
entanglement, robust.  More precisely we shall show that for any
fraction of noise there are states (and experiments to perform
upon those states) which exhibit nonlocal correlations.  The
reason that \cite{Kaszlikowski} and \cite{Durt} did not find these
experiments is because they only looked at experiments in which a
single von-Neumann measurement is made on each particle; here we
look at {\it sequences} of von-Neumann measurements.

The present discussion is, to some extent, a repeat of the history
concerning Werner's density matrices. In 1989 Werner \cite{Werner}
presented some density matrices which are entangled but which are
such that if  single von-Neumann measurements are made on each
particle, the results can be explained by a local hidden variables
model. At that time it was tacitly assumed that performing single
von-Neumann measurements on each particle essentially covers all
possibilities. However it was subsequently shown \cite{Popescu}
that the outcomes of {\it sequences} of von-Neumann measurements
are nonlocal - they cannot be explained by any hidden variables
model. This work was then extended in \cite{Popescu2},
\cite{Zukowski} and \cite{Teufel}.

We shall next explain why performing sequences of measurements
puts additional constraints on local hidden variable models, then
use this to prove that there are states with arbitrarily high
fractions of noise which exhibit nonlocal correlations.

Consider two observers, Alice and Bob, situated in two space
separated regions. The standard assumption of LHV is that if Alice
performs any arbitrary measurement $A$ and Bob performs any
arbitrary measurement $B$, and the measurements are timed so that
they take place  outside the light-cone of each other,  then there
exists a shared random variable $\lambda$, with distribution
$\mu(\lambda)$, and local distributions $P_A(a;\lambda)$ and
$P_B(b;\lambda)$ such that the joint probability that the
measurement of $A$ yields $a$ and the measurement of $B$ yields
$b$ is given by
\begin{equation}
P_{AB}(a,b)= \int P_A(a;\lambda) P_B(b;\lambda) \mu(\lambda)
d\lambda.
\end{equation}

Consider now that Alice and Bob, instead of subjecting their
particles to a single measurement, perform two measurements one
after the other, say $A_1$ followed by $A_2$  and $B_1$ followed
by $B_2$. Then a LHV model implies that
\begin{eqnarray}
\lefteqn{P_{A_1A_2B_1B_2}(a_1,a_2,b_1,b_2) =} \nonumber \\ & &
\int P_{A_1 A_2} (a_1, a_2 ;\lambda) P_{B_1 B_2}(b_1,b_2;\lambda)
\mu(\lambda) d\lambda.\label{probability}
\end{eqnarray}

Quantum mechanically the two measurements on each side could be
viewed as a single POVM. For LHV models however, doing one
measurement after the other gives us the extra constraint that we
must be able to write $P_{A_1A_2}(a_1,a_2;\lambda)$ in the form

\begin{equation}
P_{A_1A_2}(a_1,a_2;\lambda) = P_{A_1}(a_1;\lambda)
P_{A_2}(a_2;A_1, a_1, \lambda)\label{secondmeasurements}.
\end{equation} Here $P_{A_1}(a_1;\lambda)$ is the probability that Alice's particle
yields the answer $a_1$ when the first measurement to which is
subjected is $A_1$ and given that the hidden variable has the
value $\lambda$. $P_{A_2}(a_2;A_1, a_1, \lambda)$ is the
probability that Alice's particle yields the outcome $a_2$ when
the second measurement is $A_2$, given that the hidden variable
has the value $\lambda$ and given that it was first subjected to a
measurement of $A_1$ to which it yielded the outcome $a_1$. The
reason is that when Alice's particle has to give the outcome of
measurement $A_1$, it does not yet know what exactly will be the
measurement $A_2$ that will be subsequently performed, and so
cannot use that information to decide which outcome $a_1$ to give.
We must write Bob's probabilities in a similar way.

Now, let us look at the probabilities of outcomes of the second
measurement, conditioned on some fixed result of the first.
\begin{equation}
P_{A_2 B_2}(a_2, b_2; A_1, a_1, B_1, b_1) =
\frac{P_{A_1A_2B_1B_2}(a_1,a_2,b_1,b_2)}{P_{A_1B_1}(a_1b_1)}\label{conditional}.
\end{equation}

Substituting (\ref{probability}) and (\ref{secondmeasurements})
into (\ref{conditional}), and defining

\begin{equation}
\tilde{\mu}(\lambda)=\frac{  P_{A_1}(a_1;\lambda)
P_{B_1}(b_1;\lambda) } {\int P_{A_1}(a_1;\lambda)
P_{B_1}(b_1;\lambda) \mu(\lambda) d\lambda},
\end{equation}

we have that

\begin{eqnarray}
\lefteqn{P_{A_2 B_2}(a_2, b_2; A_1, a_1, B_1, b_1) =} \nonumber \\
& & \int P_{A_2}(a_2;A_1, a_1, \lambda) P_{B_2}(b_2;B_1, b_1,
\lambda)  \tilde{\mu} (\lambda) d\lambda.
\end{eqnarray}

We shall now only consider experiments in which the first
measurements are fixed and give some particular fixed outcomes,
and thus can drop the indices $A_1$, $a_1$, $B_1$ and $b_1$, which
leaves us with
\begin{equation}
P_{A_2B_2}(a_2b_2) = \int P_{A_2}(a_2; \lambda) P_{B_2}(b_2;
\lambda) \tilde{\mu}(\lambda) d \lambda\label{post-selectedlhv}.
\end{equation} We further note that $\tilde{\mu}(\lambda)$ is
positive and $\int \tilde{\mu}(\lambda) d \lambda=1$, thus it can
be viewed as a probability distribution analogously to
$\mu(\lambda)$. Thus, if the whole experiment could be explained
by a local hidden variables model, then the probabilities of
 outcomes for the second measurement conditioned upon
any result of the first measurement have to be given by a LHV
model themselves.  This is a consequence of doing the measurements
one after the other rather than together.  In particular, we can
look at Bell inequalities for these conditioned probabilities, and
know that if they are violated, then the initial state is
nonlocal. For example suppose that the second measurement which is
performed by Alice is either $A_2$ or $A_2'$ and that performed by
Bob is either $B_2$ or $B_2'$. Then using the CHSH inequality
\cite{CHSH} (a particular Bell type inequality) and
(\ref{post-selectedlhv}) it follows that
\begin{equation}
E(A_2B_2)+E(A_2B_2')+E(A_2'B_2)-E(A_2'B_2')\leq 2\label{CHSH}.
\end{equation}
Here $E(A_2B_2)=Tr\tilde\rho A_2B_2$ is the expectation value of
the product of the operators $A_2$ and $B_2$ in the state
$\tilde\rho$ which is the state of the system after the first
measurements (assuming that we indeed obtained the particular
fixed outcomes we have chosen).

We shall now use (\ref{CHSH}) to show that for sufficiently large
$N$, the states defined in equation (\ref{state}) generate
nonlocal correlations. We take the first measurement on Alice's
side, $A_1$, to be the projection onto the subspace
$\{|1>_A,|2>_A\}$. The first measurement on Bob's side, $B_1$, is
the projection onto the subspace $\{|1>_B,|2>_B\}$.   We just look
at the cases where the state is indeed in the first two subspaces,
in which case the state becomes (after the first measurements):

\begin{equation}
\tilde\rho=\frac{(1- F_N)N}{N(1-F_N)+2F_N} \ket{\Psi_2}
\bra{\Psi_2} +
 \frac{2F_N}{N(1-F_N)+2F_N} \frac{\hat{I}_{2 \times 2}}{2^2}.
\end{equation}

We now take the second measurements ($A_2$, $A_2'$, $B_2$, $B_2'$)
to be those which give the maximal violation of the CHSH
inequality on the state $\ket{\Psi_2}_{AB}$, and we note that if
the CHSH inequality is violated, the initial state is nonlocal.
This occurs when
\begin{equation}
F_N < \frac{N}{N+c},
\end{equation}
where $c= \frac{2}{\sqrt{2}-1} \approx 4.83$. Therefore, for any
fraction of noise we can, by taking $N$ large enough, find states
which give nonlocal correlations. Thus we have shown that the
nonlocal correlations are robust to noise.

Finally, we note that we have not completely solved the problem of
which states of the form (\ref{state}) generate nonlocal
correlations. Recalling that [7-9] states of this form are
separable iff $F_N \ge \frac{N}{N+1}$, we can see that the states
for which $\frac{N}{N+c} \le F_N < \frac{N}{N+1}$ are entangled
but do not violate the Bell inequality we have considered. It is
an interesting and open question as to whether these states
generate nonlocal correlations or not.

 \end{multicols}
\end{document}